\documentclass[%
 reprint,
 amsmath,amssymb,
 aps,
]{revtex4-2}

\usepackage{graphicx}
\usepackage{dcolumn}
\usepackage{bm}

\begin{document}

\title{High-Precision Acoustic Event Monitoring in Single-Mode Fibers Using Fisher Information}

\author{Catarina S. Monteiro}
 \homepage{catarina.s.monteiro@inesctec.pt}
\author{Tiago D. Ferreira}
\author{Nuno A. Silva}

\affiliation{INESC TEC - Institute for Systems and Computer
Engineering, Technology and Science, Rua do Campo Alegre 687, 4169-007 Porto, Portugal}
\affiliation{Departamento de Física e Astronomia, Faculdade de Ciências, Universidade do Porto, Rua do Campo Alegre s/n, 4169-007 Porto, Portugal}

\date{\today}

\begin{abstract}
Polarization optical fiber sensors are based on modifications of fiber birefringence by an external measurand (e.g. strain, pressure, acoustic waves). Yet, this means that different input states of polarization will result in very distinct behaviors, which may or may not be optimal in terms of sensitivity and signal-to-noise ratio. To tackle this challenge, this manuscript presents an optimization technique for the input polarization state using the Fisher information formalism, which allows for achieving maximal precision for a statistically unbiased metric. By first measuring the variation of the Mueller matrix of the optical fiber in response to controlled acoustic perturbations induced by piezo speakers, we compute the corresponding Fisher information operator. Using maximal information states of the Fisher information, it was possible to observe a significant improvement in the performance of the sensor, increasing the signal-to-noise ratio from 4.3 to 37.6 dB, attaining an almost flat response from 1.5 kHz up to 15 kHz. As a proof-of-concept for dynamic audio signal detection, a broadband acoustic signal was also reconstructed with significant gain, demonstrating the usefulness of the introduced formalism for high-precision sensing with polarimetric fiber sensors.

\end{abstract}

\maketitle

\section{Introduction}

Polarization-based optical fiber sensors exploit birefringence-induced changes in the state of polarization (SoP) of light to measure physical parameters such as strain, temperature, and pressure with high resolution. These changes are typically quantified using the Stokes formalism and interrogated using polarimetric techniques such as Mueller matrix analysis or polarization-resolved reflectometry \cite{Tong2015, Tang2024}. These methods allow for distributed and high-speed interrogation and are widely applied in structural health monitoring, geophysical sensing, and industrial diagnostics
\cite{Naeem2015, Muller2019, Palmieri2015, Carver2024}. 

Despite these advantages, achieving optimal sensitivity and estimation precision is challenging, as each possible input SoP drives a different response. As a result, the sensitivity and even the shape of the transfer function depend strongly on the input state. Besides, issues such as polarization modal dispersion, modal coupling effects, and mechanical instability can also degrade repeatability, signal-to-noise ratio (SNR), and long-range sensing accuracy \cite{Palmieri2013}. All of these mean that precise polarization control of the input SoP is crucial. Polarization management solutions typically rely on mechanically rotated wave plates of optical fiber coils \cite{JoonYongCho2005, Wang2015}, which, despite their simplicity, suffer from limited speed and are susceptible to mechanical vibrations that can introduce instability \cite{Lin2022}. For higher speed polarization management, it has been reported the use of liquid crystal devices \cite{Sun2022, Park2025}, hybrid systems combining waveplates and magneto-optic crystals \cite{Zhang2006, Stoyanova2020}, or piezo-actuated fibers \cite{Li2007, Meng2022}, which however increase cost and calibration overhead while offering no guarantee that the chosen SOP is statistically optimal. Both these solutions can be utilized to seek optimal operation points or average out the dependence, yet a statistically optimal method for selecting the launch SOP is still missing.  In parallel, data-driven post-processing \cite{Awad2024, Usmani2025} can rescue some performance but limits their adaptability across different sensing scenarios and hinders their ability to achieve the theoretical bounds of precision and sensitivity.

To address this gap, we propose a novel approach grounded on the Fisher information formalism, a statistical framework for quantifying the amount of information that a measured variable $X$ carries about an unknown parameter $\theta$ \cite{Djordjevic2022}. By modeling and maximizing the Fisher information, it is possible to achieve the optimal operation conditions, with the advantage of obtaining a fundamental lower bound on the precision of unbiased estimators called the Cramér-Rao bound. This concept has been explored in a wide range of fields such as quantum metrology \cite{Lu2015, Du2025}, machine learning \cite{Hannun2022}, coherent wavefront shaping \cite{Bouchet2021,Gutierrez-Cuevas2024}, and communications \cite{Shen2007, Barnes2019}, but in the context of optical fiber sensing it remains largely unexplored outside the quantum sensing field\cite{Lee2022, Peng2023}. In the particular case of polarization-based sensors, the Fisher information may play the role of a powerful agnostic tool for optimizing sensor configurations by identifying the polarization states that maximize sensitivity to specific external perturbations, thereby improving estimation precision without relying on hardware complexity or empirical calibration.

In this work, we explore a statistically rigorous Fisher information framework to obtain performance gains for an optical fiber acoustic sensor based on SoP measurements. By characterizing the perturbation-induced changes on the Mueller matrix and constructing a Fisher operator, we show that it is possible to identify the input polarization state that maximizes the sensitivity to an applied external stimulus. Comparing optimal with non-optimal SoP, the results presented show a nine-fold improvement in the signal-to-noise ratio (SNR), from 4.3 to 37.6~dB, or an almost 40 times increase in amplitude-based sensitivity and a nearly flat frequency response for frequencies from 1.5~kHz to 15~kHz. Finally, as a proof-of-concept of the method applied to polarimetric-based acoustic sensing, we accurately reconstruct a complex audio stimulus, confirming enhanced precision and broadband interrogation, and showcasing the potential of this interrogation technique for high-precision, real-time sensing using a statistically rigorous and hardware-efficient methodology.

\section{Modeling polarization evolution and Fisher information-based optimization}

In order to introduce a Fisher information formalism, one first needs to introduce a mathematical model to relate the measured output with the input SoP. The state of polarization of light may be fully described by the Stokes vector,  $\mathbf{S} = \left[ ~S_0, ~S_1, ~S_2, ~S_3\right]^{T}$, where $S_0$, $S_1$, $S_2$, $S_3$ are observables of the polarized field. Specifically, $S_0$ denotes the total intensity of the light; $S_1$ quantifies the preponderance of vertical over the horizontal linear polarization; $S_2$ represents the preponderance of linear polarization at +45º over -45º; and $S_3$ describes the predominance of right over left-handed circular polarization. 

As the light propagates through a birefringent medium such as an optical fiber, its polarization state evolves due to intrinsic and extrinsic perturbations. Mathematically, this evolution can be characterized by the Mueller matrix $\bar{M}$, which governs the transformation of the Stokes vector by setting a relationship between the input Stokes vector $\mathbf{S}_{in}$ and their corresponding output Stokes vector $\mathbf{S}_{out}$ given by \cite{Gao2021}:

\begin{equation}
    \mathbf{S}_{out} = \bar{M} \cdot \mathbf{S}_{in}.
\end{equation}

Assuming normalized intensity and focusing on the polarization components only, the Stokes vector can be reduced to $\mathbf{S} = [S_1, S_2, S_3]^T$, and the corresponding Mueller matrix reduced to a 3$\times$3 matrix.. 

\subsection{Sensing, estimation, and optimization via Fisher information}

The intrinsic birefringence of the optical fiber is sensitive to external perturbations such as axial strain, temperature variations, or local curvature, altering the local refractive index tensor and, thus, modifying the polarization state of the traveling light. The polarization evolution, governed by the Mueller matrix $\bar{M}(\zeta)$, becomes a function of the applied stimulus $\zeta$. For sufficiently small perturbations, a linear model comes in the form of a first-order Taylor expansion, enabling the perturbation to be estimated directly by the measured change in the output Stokes vector ($\mathbf{S}_{out}$) as:
\begin{equation}
    \zeta = \frac{\left (\mathbf{S}_{out} - 
    \mathbf{S}_{out}^{(\zeta = 0)} \right ) \cdot \left ( \partial_\zeta \bar{M} \cdot \mathbf{S}_{in} \right )^\dagger }{ \left \| \partial_\zeta \bar{M}\cdot \mathbf{S}_{in} \right \|^2}
\label{eq:estimation}
\end{equation}
where $\mathbf{S_{out}}^{(\zeta = 0)}$ is the baseline Stokes vector with no perturbation, $\partial_\zeta \bar{M} $ is the partial derivative of the Mueller matrix with respect to $\zeta$, approximated experimentally using a finite difference method. From this expression, it becomes evident that the precision and sensitivity of the perturbation estimation are intrinsically dependent on the input polarization state $\mathbf{S}_{in}$. It can be shown that expression \ref{eq:estimation} corresponds to the minimum variance unbiased estimator and that, in such conditions, the Fisher information formalism can be used to maximize the performance of the system\cite{Bouchet2021,Gutierrez-Cuevas2024}.

For this case, the Fisher information can be calculated as $\mathcal{J} = \mathbb{E} ([\partial_\zeta ln~p(\mathbf{S}_{out}; \zeta)]^2)$ with $p(X; \theta)$ standing for the probability density function of the observed data $\mathbf{S}_{out}$ given $\zeta$, and $\mathbb{E}$ being the expected value acting over noise fluctuations. Formally speaking, it quantifies the amount of information an observable random variable $\mathbf{S}_{out}$ carries about an unknown parameter $\zeta$, and is directly linked to the Cramér-Rao inequality that establishes a lower bound on the variance ${\sigma}^2_\zeta$ of any unbiased estimator as ${\sigma}^2_\zeta \geqslant \mathcal{J}^{-1}$. 

Assuming the noise in each of the Stokes parameters to be Gaussian, with known standard deviation $\sigma$, and that each parameter is a statistically independent variable (i.e. $\sigma$ is sufficiently small), the Fisher information can be written in terms of the input Stokes vector $\mathbf{S}_{in}$ as \cite{Bouchet2021,Gutierrez-Cuevas2024}
\begin{equation}
   \mathcal{J}(\theta) = \frac{1}{\sigma^2}\langle{\mathbf{S}_{in}}|{\bar{F}_{\zeta}}|{\mathbf{S}_{in}}\rangle
   \label{eq:FisherInformation}
\end{equation}
where bra-ket notation represents the complex inner product and the Fisher operator $\bar{F}_{\zeta}$ given by
\begin{equation}
        \bar{F}_{\zeta} = (\partial_\zeta \bar{M})^\dagger \partial_\zeta \bar{M}
    \label{eq:FisherInformationOperator}
\end{equation}
with $\dagger$ standing for the conjugate transpose. As the operator $\bar{F}_\zeta$ is Hermitian, the eigenvalues are real and non-negative, and its principal eigenvector corresponds to the input polarization state that maximizes Fisher information. The associated largest eigenvalue then defines the maximum achievable information content for the system concerning perturbation $\zeta$.

\section{Methods and Results}
\subsection{Experimental Setup}
\begin{figure}[ht!]
\centering\includegraphics[width=0.45\textwidth]{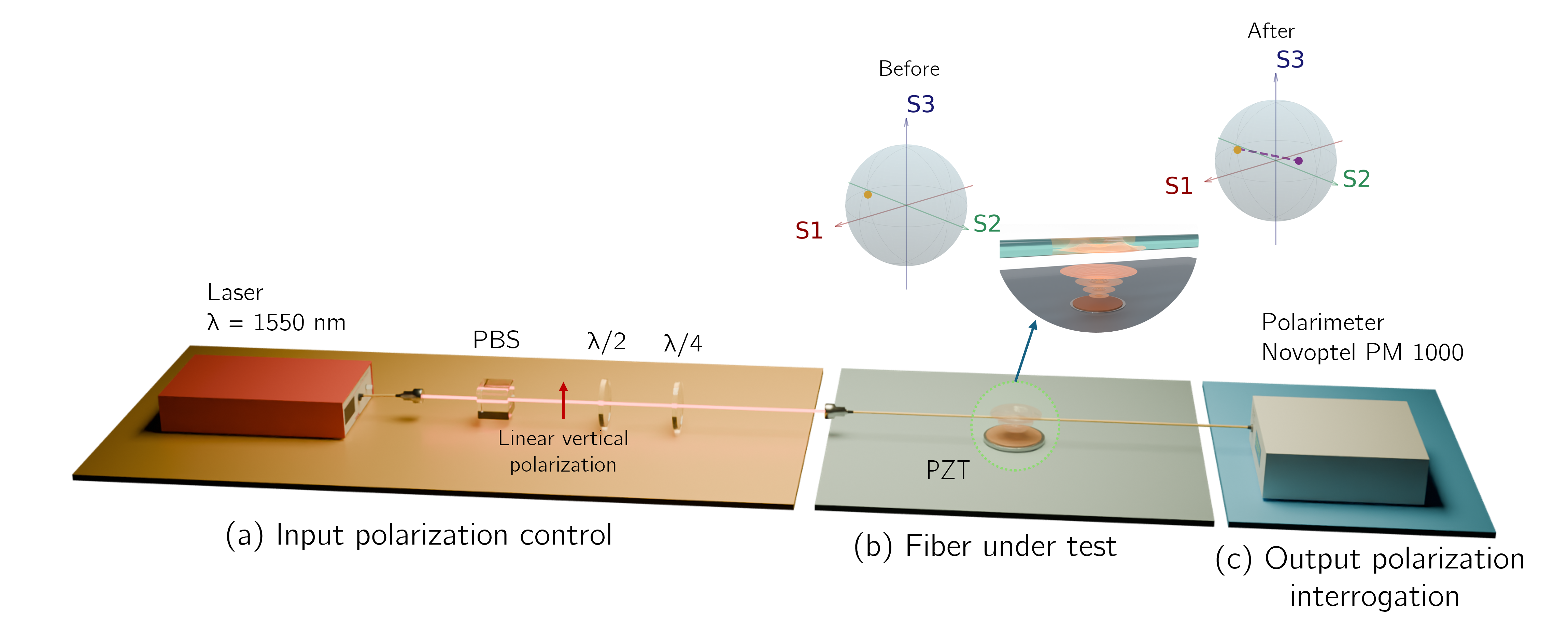}
\caption{Experimental setup composed by (a) input polarization control - a collimated tunable laser with operating wavelength set at 1550 nm that passes through a PBS guarantying a linear vertically polarized input light, a half-wave plate and a quarter-wave plate, allowing a full-control of the input polarization state; (b) fiber under test - an SMF with an approximate length of 3~m is subjected to acoustic perturbations using piezoelectric actuators driven by a DAQ system (omitted from the figure); (c) output polarization interrogation carried out by a high-speed polarimeter.}
\label{img-setup}
\end{figure}

To probe the influence of external perturbations on the polarization state of guided light, an experimental setup comprising a laser, polarization control optics, a single-mode optical fiber, and a high-speed polarimetric detection system was employed as schematized in Figure \ref{img-setup}. On the polarization control stage, a laser (Thorlabs TL200C), operating at a wavelength 1550~nm is collimated before entering the polarization control optics composed of a polarized beam splitter (PBS), a half-wave plate and a quarter-wave plate mounted on motorized rotation stages (Thorlabs ELL14), which allow precise, fast, and automatic control of the input polarization state before coupling the light to the fiber under test with a second collimator. The fiber - a standard single-mode fiber (SMF-28), with an approximate total length of 3~m - is then subjected to external acoustic perturbations using a piezoelectric actuator, which is driven by computer-generated signals via a data acquisition system (NI 6229) operating at a sampling rate of 100 kHz. At the output, the polarization state is measured using a high-speed polarimeter (Novoptel PM1000), enabling real-time acquisition of the Stokes parameters at a maximum sampling rate of 100 MHz.

\subsection{Mueller Matrix determination}
The Mueller matrix $\bar{M}$ can be calculated experimentally by measuring the output Stokes vectors corresponding to three orthogonal input polarization states  $\mathbf{S}_{1}^{in} = [1, 0, 0]^T$, $\mathbf{S}_{2}^{in} = [0, 1, 0]^T$, and $\mathbf{S}_{3}^{in} = [0, 0, 1]^T$. The resulting output vectors, $\mathbf{S}_i^{out}$, form the columns of the matrix $\bar{M}$, enabling full characterization of the polarization transformation that can be experimentally validated via matrix inversion\cite{Popoff2010}.

\subsection{Acoustic signal detection optimization via Fisher information}
Acoustic signals that reach the optical fiber will induce local perturbations in the birefringence of the fiber, creating dynamic changes in the output Stokes vector. However, these changes are often masked by system noise or limited by the intrinsic sensitivity of the polarization state of the propagating light. Under the context of this work, the hypothesis is that computing the Fisher information of the system allows to identify the configuration that maximizes the amount of information extracted about the acoustic-induced deformation. 

To experimentally determine the Fisher operator defined in equation \ref{eq:FisherInformationOperator}, the Mueller matrix was measured under both unperturbed and perturbed conditions. This was accomplished by applying a sinusoidal acoustic signal to the piezoelectric actuator with a frequency of 4~kHz and a peak amplitude with drive level of the actuator of 2~V. This configuration enabled the acquisition of the unperturbed Mueller matrix $\bar{M}_0$ and perturbed one $\bar{M}_{\zeta}$, from which the derivative $\partial_{\zeta} \bar{M} $ was estimated using a first-order finite difference approach. 

\begin{figure}[ht!]
\centering\includegraphics[width=0.45\textwidth]{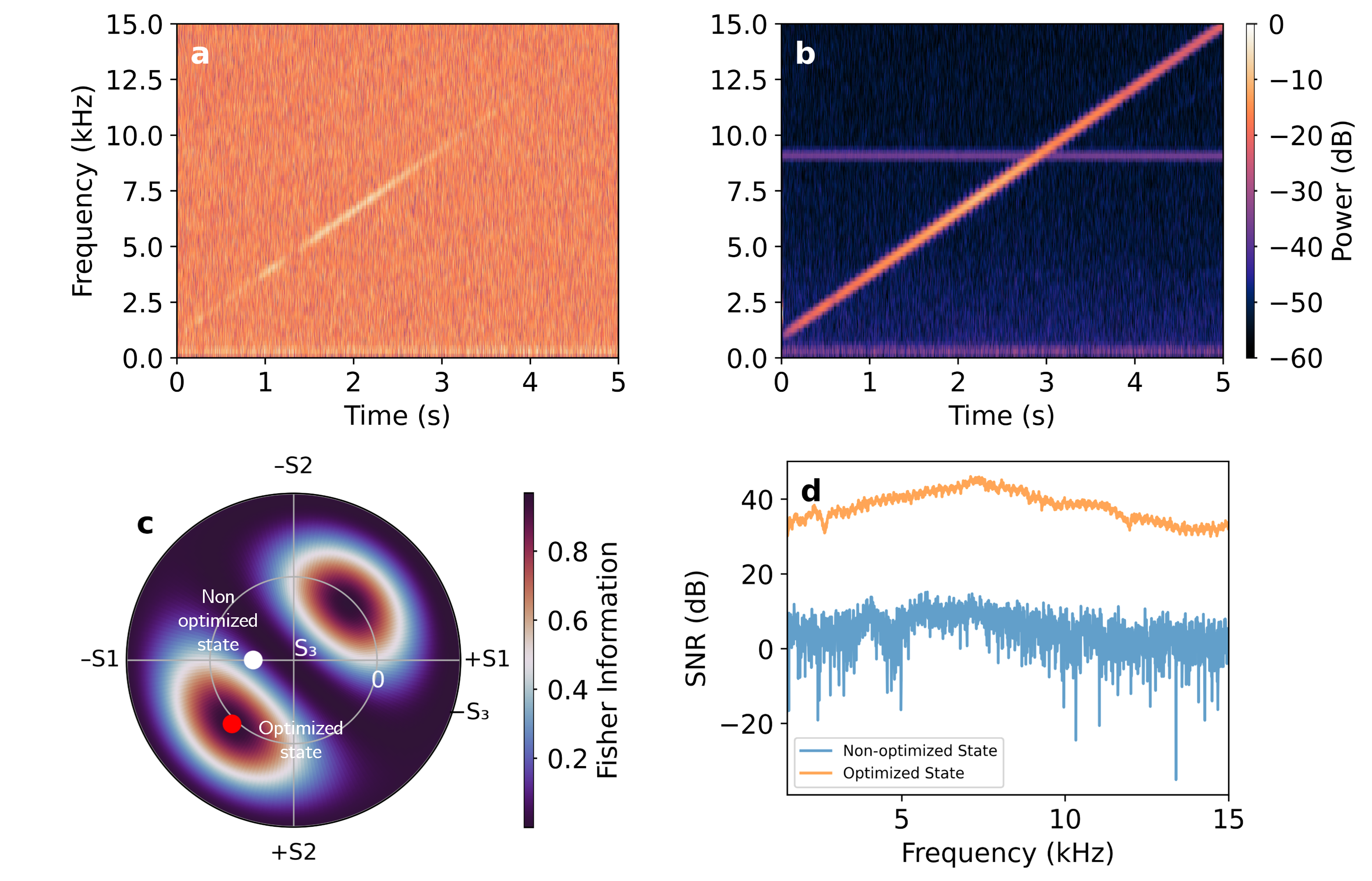}
\caption{Spectrograms of the perturbation estimate, computed with the short-time Fourier transform, for (a) a reference input polarization state with Stokes vector $\mathbf{S} = [~ -0.684,~ -0.007,~  0.729~]$ and (b) the Fisher-optimized state that corresponds to the maximum information state. (c) Polar‐map representation of the Poincaré sphere, with the color scale indicating the magnitude of the Fisher information with with the non-optimized and the optimized states annotated in white and red, correspondingly. Panel (d) plots the frequency-resolved signal-to-noise ratio obtained with the two states; the Fisher-optimized state yields an almost flat response and increases the mean SNR from 4.3 to 37.6 dB. }
\label{img-MuellerMatrix_spectrogram}
\end{figure}

Following system calibration, the Fisher information was computed for a range of input states using equation \ref{eq:FisherInformation}, and the optimized state was calculated using singular value decomposition (SVD) to get the most sensitive input polarization state (see Poincaré Sphere in Figure \ref{img-MuellerMatrix_spectrogram}(c)). To assess the system's performance and the improvement achieved using Fisher-improved input state, a chirped signal with frequencies between 1.5 and 15 kHz and a fixed peak amplitude of 0.2~V was applied to the fiber under different polarization input states. Figure \ref{img-MuellerMatrix_spectrogram}, show the spectrograms of the estimated perturbations $\zeta$, calculated using equation \ref{eq:estimation}, (a) a Stokes input state $\mathbf{S} = [~ -0.684,~ -0.007,~  0.729~]$, (b) the state yielding the maximum Fisher information, and (d) the frequency-resolved signal-to-noise ratio (SNR) for the two states. The SNR was calculated at each instant $i$ as the difference between the peak spectrogram power within $\pm$150~Hz of the instantaneous sweep frequency and the median power outside that band, in dB. It was calculated as $SNR(i) [dB] = max_{|f-f(i)| \le \Delta f} S(f,t_i) - median_{|f-f(i)| > \Delta f} S(f,t_i)$, where $i$ indexes the time-frame, $S$ the spectrogram power in dB, and $\Delta f$ represents the half-bandwidth of the tolerance around the instantaneous frequency. As observed, the application of the Fisher-optimal input state yielded a measurable improvement in the SNR, increasing from an average 4.3~dB to an average 37.6~dB for the Fisher-optimized state.

\subsection{Quantitative assessment of signal fidelity}
\begin{figure}[ht!]
\centering\includegraphics[width=0.45\textwidth]{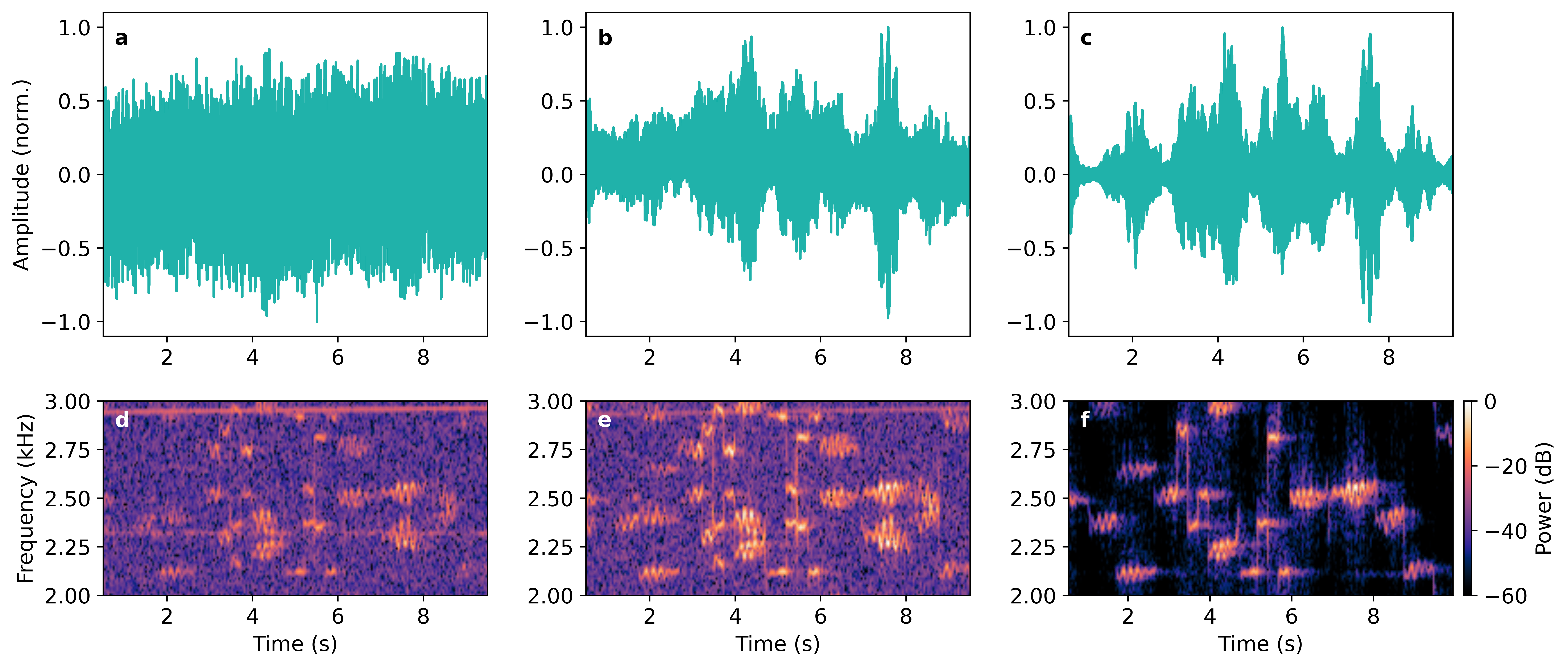}
\caption{Time-domain and time-frequency responses of the perturbation estimates under broadband acoustic excitation. (a) Estimated perturbation signal using a reference input polarization state $\mathbf{S}_{ref} = [~ -0.684,~ -0.007,~  0.729~]$; (b) estimated signal using the Fisher-optimized input polarization state; (c) applied audio waveform; (d-e) spectrograms of the estimated signals in (a) and (b), respectively, showing improved temporal and spectral resolution under Fisher optimization; (f) spectrogram of the input signal, used as a reference for comparison. }
\label{img-dynamic_spectrogram}
\end{figure}

The robustness of the Fisher-optimized interrogation was assessed by applying a broadband, dynamically modulated excitation, using excerpts of orchestral music with a sampling rate of 48 kHz and a 0-3 kHz bandwidth. The amplitude of the audio waveform was normalized to a peak drive of 0.1 V and applied to the piezoelectric transducer. Using the previously calibrated Mueller‑matrix derivatives, the time‑resolved Stokes vector was recorded, and the perturbation estimates were reconstructed using both a reference polarization state and the Fisher-optimized state. Qualitatively, the time-domain response Figure \ref{img-dynamic_spectrogram}(a,b) and spectrograms Figure \ref{img-dynamic_spectrogram} (d,e) show that the Fisher-optimized input state significantly enhances the ability of the system to track and reconstruct complex, time-varying acoustic signals. Quantitatively, using root mean squared error (RMSE) between the estimated perturbation signal and the original audio waveform as a metric, it decreased from 0.40 (non-optimal state) to 0.31 (Fisher-optimized state), reflecting a 22.5\% reduction in the estimation error of the applied waveform.

\section{Conclusions}
This work explored the use of Fisher information as a metric to find the optimal operation point and performance of polarization-based optical fiber sensors in standard single-mode fibers. By characterizing the Mueller matrix response of the fiber under controlled acoustic perturbations, we demonstrated that the optimal input polarization state can be identified, maximizing the sensitivity of the sensor. In particular, experimental results demonstrate a substantial increase in measurement quality, with a signal-to-noise ratio improving from 4.3~dB (non-optimal random state) up to 37.6~dB, representing a nearly 40-fold increase in amplitude-based sensitivity with a nearly flat frequency response covering the audible spectrum, from 1.5 kHz to 15 kHz. Focusing on the reconstruction of the waveform itself in a more complex and real-world scenario (e.g., music reconstruction), the Fisher-optimized state presented an improvement on the root mean squared error (RMSE) from 0.40 in the reference state to 0.31, reflecting a 22.5\% reduction in the estimated error.

Overall, the findings enclosed demonstrate the potential of the proposed methodology for an expedite identification of the optimal input polarization state, eliminating the need for manual scanning or trial-and-error calibration. Providing a statistically grounded method to determine the input state that offers greatest information content, it also establishes an interesting connection with the theoretical precision limit, a relation that can be further explored in future works and contexts closer to ultimate precision such as quantum sensing. Besides, being hardware and stimuli agnostic, it is ready to be integrated into existing polarimetric systems, paving also a path for real-time polarimetric sensing in distributed fiber networks, multi-parameter estimation, and adaptive sensing strategies driven by feedback or machine learning, in particular, in scenarios of low signal-to-noise ratio.

\begin{acknowledgments}
This work is co-financed by Component 5 - Capitalization and Business Innovation, integrated in the Resilience Dimension of the Recovery and Resilience Plan within the scope of the Recovery and Resilience Mechanism (MRR) of the European Union (EU), framed in the Next Generation EU, for the period 2021 - 2026, within project HfPT, with
reference 41, and by national funds through FCT – Fundação para a Ciência e a Tecnologia, I.P., under the support UID/50014/2023 (https://doi.org/10.54499/UID/50014/2023). Nuno A. Silva acknowledges the support of FCT under the grant 2022.08078.CEECIND/CP1740/CT0002 (https://doi.org/10.54499/2022.08078.CEECIND/CP1740/CT0002).

The authors also gratefully acknowledge Orlando Frazão and Susana Silva for providing access to experimental equipment.

\end{acknowledgments}

\bigskip


\bibliography{sample}

\begin{thebibliography}{30}%
\makeatletter
\providecommand \@ifxundefined [1]{%
 \@ifx{#1\undefined}
}%
\providecommand \@ifnum [1]{%
 \ifnum #1\expandafter \@firstoftwo
 \else \expandafter \@secondoftwo
 \fi
}%
\providecommand \@ifx [1]{%
 \ifx #1\expandafter \@firstoftwo
 \else \expandafter \@secondoftwo
 \fi
}%
\providecommand \natexlab [1]{#1}%
\providecommand \enquote  [1]{``#1''}%
\providecommand \bibnamefont  [1]{#1}%
\providecommand \bibfnamefont [1]{#1}%
\providecommand \citenamefont [1]{#1}%
\providecommand \href@noop [0]{\@secondoftwo}%
\providecommand \href [0]{\begingroup \@sanitize@url \@href}%
\providecommand \@href[1]{\@@startlink{#1}\@@href}%
\providecommand \@@href[1]{\endgroup#1\@@endlink}%
\providecommand \@sanitize@url [0]{\catcode `\\12\catcode `\$12\catcode `\&12\catcode `\#12\catcode `\^12\catcode `\_12\catcode `\%12\relax}%
\providecommand \@@startlink[1]{}%
\providecommand \@@endlink[0]{}%
\providecommand \url  [0]{\begingroup\@sanitize@url \@url }%
\providecommand \@url [1]{\endgroup\@href {#1}{\urlprefix }}%
\providecommand \urlprefix  [0]{URL }%
\providecommand \Eprint [0]{\href }%
\providecommand \doibase [0]{https://doi.org/}%
\providecommand \selectlanguage [0]{\@gobble}%
\providecommand \bibinfo  [0]{\@secondoftwo}%
\providecommand \bibfield  [0]{\@secondoftwo}%
\providecommand \translation [1]{[#1]}%
\providecommand \BibitemOpen [0]{}%
\providecommand \bibitemStop [0]{}%
\providecommand \bibitemNoStop [0]{.\EOS\space}%
\providecommand \EOS [0]{\spacefactor3000\relax}%
\providecommand \BibitemShut  [1]{\csname bibitem#1\endcsname}%
\let\auto@bib@innerbib\@empty
\bibitem [{\citenamefont {Tong}\ \emph {et~al.}(2015)\citenamefont {Tong}, \citenamefont {Dong}, \citenamefont {Wang}, \citenamefont {Sun}, \citenamefont {Wang}, \citenamefont {Bai}, \citenamefont {Yuan}, \citenamefont {Zhu},\ and\ \citenamefont {Liu}}]{Tong2015}%
  \BibitemOpen
  \bibfield  {author} {\bibinfo {author} {\bibfnamefont {Y.}~\bibnamefont {Tong}}, \bibinfo {author} {\bibfnamefont {H.}~\bibnamefont {Dong}}, \bibinfo {author} {\bibfnamefont {Y.}~\bibnamefont {Wang}}, \bibinfo {author} {\bibfnamefont {W.}~\bibnamefont {Sun}}, \bibinfo {author} {\bibfnamefont {X.}~\bibnamefont {Wang}}, \bibinfo {author} {\bibfnamefont {J.}~\bibnamefont {Bai}}, \bibinfo {author} {\bibfnamefont {H.}~\bibnamefont {Yuan}}, \bibinfo {author} {\bibfnamefont {N.}~\bibnamefont {Zhu}},\ and\ \bibinfo {author} {\bibfnamefont {J.}~\bibnamefont {Liu}},\ }\bibfield  {title} {\bibinfo {title} {{Distributed incomplete polarization-OTDR based on polarization maintaining fiber for multi-event detection}},\ }\href {https://doi.org/10.1016/j.optcom.2015.06.057} {\bibfield  {journal} {\bibinfo  {journal} {Optics Communications}\ }\textbf {\bibinfo {volume} {357}},\ \bibinfo {pages} {41} (\bibinfo {year} {2015})}\BibitemShut {NoStop}%
\bibitem [{\citenamefont {Tang}\ \emph {et~al.}(2024)\citenamefont {Tang}, \citenamefont {Zhu}, \citenamefont {Pang}, \citenamefont {Wei}, \citenamefont {Zhang}, \citenamefont {Chen},\ and\ \citenamefont {Wang}}]{Tang2024}%
  \BibitemOpen
  \bibfield  {author} {\bibinfo {author} {\bibfnamefont {Y.}~\bibnamefont {Tang}}, \bibinfo {author} {\bibfnamefont {M.}~\bibnamefont {Zhu}}, \bibinfo {author} {\bibfnamefont {F.}~\bibnamefont {Pang}}, \bibinfo {author} {\bibfnamefont {H.}~\bibnamefont {Wei}}, \bibinfo {author} {\bibfnamefont {L.}~\bibnamefont {Zhang}}, \bibinfo {author} {\bibfnamefont {W.}~\bibnamefont {Chen}},\ and\ \bibinfo {author} {\bibfnamefont {T.}~\bibnamefont {Wang}},\ }\bibfield  {title} {\bibinfo {title} {{Distributed optical fiber magnetic field sensor based on polarization-sensitive OFDR}},\ }\href {https://doi.org/10.1364/oe.511557} {\bibfield  {journal} {\bibinfo  {journal} {Optics Express}\ }\textbf {\bibinfo {volume} {32}},\ \bibinfo {pages} {11726} (\bibinfo {year} {2024})}\BibitemShut {NoStop}%
\bibitem [{\citenamefont {Naeem}\ \emph {et~al.}(2015)\citenamefont {Naeem}, \citenamefont {Kim}, \citenamefont {Kim},\ and\ \citenamefont {Chung}}]{Naeem2015}%
  \BibitemOpen
  \bibfield  {author} {\bibinfo {author} {\bibfnamefont {K.}~\bibnamefont {Naeem}}, \bibinfo {author} {\bibfnamefont {B.~H.}\ \bibnamefont {Kim}}, \bibinfo {author} {\bibfnamefont {B.}~\bibnamefont {Kim}},\ and\ \bibinfo {author} {\bibfnamefont {Y.}~\bibnamefont {Chung}},\ }\bibfield  {title} {\bibinfo {title} {{Simultaneous multi-parameter measurement using Sagnac loop hybrid interferometer based on a highly birefringent photonic crystal fiber with two asymmetric cores}},\ }\href {https://doi.org/10.1364/oe.23.003589} {\bibfield  {journal} {\bibinfo  {journal} {Optics Express}\ }\textbf {\bibinfo {volume} {23}},\ \bibinfo {pages} {3589} (\bibinfo {year} {2015})}\BibitemShut {NoStop}%
\bibitem [{\citenamefont {Muller}\ \emph {et~al.}(2019)\citenamefont {Muller}, \citenamefont {Frank}, \citenamefont {Yang}, \citenamefont {Gu},\ and\ \citenamefont {Bohnert}}]{Muller2019}%
  \BibitemOpen
  \bibfield  {author} {\bibinfo {author} {\bibfnamefont {G.~M.}\ \bibnamefont {Muller}}, \bibinfo {author} {\bibfnamefont {A.}~\bibnamefont {Frank}}, \bibinfo {author} {\bibfnamefont {L.}~\bibnamefont {Yang}}, \bibinfo {author} {\bibfnamefont {X.}~\bibnamefont {Gu}},\ and\ \bibinfo {author} {\bibfnamefont {K.}~\bibnamefont {Bohnert}},\ }\bibfield  {title} {\bibinfo {title} {{Temperature Compensation of Interferometric and Polarimetric Fiber-Optic Current Sensors with Spun Highly Birefringent Fiber}},\ }\href {https://doi.org/10.1109/JLT.2019.2907803} {\bibfield  {journal} {\bibinfo  {journal} {Journal of Lightwave Technology}\ }\textbf {\bibinfo {volume} {37}},\ \bibinfo {pages} {4507} (\bibinfo {year} {2019})}\BibitemShut {NoStop}%
\bibitem [{\citenamefont {Palmieri}\ \emph {et~al.}(2015)\citenamefont {Palmieri}, \citenamefont {Sarchi},\ and\ \citenamefont {Galtarossa}}]{Palmieri2015}%
  \BibitemOpen
  \bibfield  {author} {\bibinfo {author} {\bibfnamefont {L.}~\bibnamefont {Palmieri}}, \bibinfo {author} {\bibfnamefont {D.}~\bibnamefont {Sarchi}},\ and\ \bibinfo {author} {\bibfnamefont {A.}~\bibnamefont {Galtarossa}},\ }\bibfield  {title} {\bibinfo {title} {{Distributed measurement of high electric current by means of polarimetric optical fiber sensor}},\ }\href {https://doi.org/10.1364/oe.23.011073} {\bibfield  {journal} {\bibinfo  {journal} {Optics Express}\ }\textbf {\bibinfo {volume} {23}},\ \bibinfo {pages} {11073} (\bibinfo {year} {2015})}\BibitemShut {NoStop}%
\bibitem [{\citenamefont {Carver}\ and\ \citenamefont {Zhou}(2024)}]{Carver2024}%
  \BibitemOpen
  \bibfield  {author} {\bibinfo {author} {\bibfnamefont {C.~J.}\ \bibnamefont {Carver}}\ and\ \bibinfo {author} {\bibfnamefont {X.}~\bibnamefont {Zhou}},\ }\bibfield  {title} {\bibinfo {title} {{Polarization sensing of network health and seismic activity over a live terrestrial fiber-optic cable}},\ }\href {https://doi.org/10.1038/s44172-024-00237-w} {\bibfield  {journal} {\bibinfo  {journal} {Communications Engineering}\ }\textbf {\bibinfo {volume} {3}},\ \bibinfo {pages} {1} (\bibinfo {year} {2024})}\BibitemShut {NoStop}%
\bibitem [{\citenamefont {Palmieri}(2013)}]{Palmieri2013}%
  \BibitemOpen
  \bibfield  {author} {\bibinfo {author} {\bibfnamefont {L.}~\bibnamefont {Palmieri}},\ }\bibfield  {title} {\bibinfo {title} {{Distributed polarimetric measurements for optical fiber sensing}},\ }\href {https://doi.org/10.1016/j.yofte.2013.07.015} {\bibfield  {journal} {\bibinfo  {journal} {Optical Fiber Technology}\ }\textbf {\bibinfo {volume} {19}},\ \bibinfo {pages} {720} (\bibinfo {year} {2013})}\BibitemShut {NoStop}%
\bibitem [{\citenamefont {{Joon Yong Cho}}\ \emph {et~al.}(2005)\citenamefont {{Joon Yong Cho}}, \citenamefont {{Jong Hoon Lim}},\ and\ \citenamefont {{Kyung Shik Lee}}}]{JoonYongCho2005}%
  \BibitemOpen
  \bibfield  {author} {\bibinfo {author} {\bibnamefont {{Joon Yong Cho}}}, \bibinfo {author} {\bibnamefont {{Jong Hoon Lim}}},\ and\ \bibinfo {author} {\bibnamefont {{Kyung Shik Lee}}},\ }\bibfield  {title} {\bibinfo {title} {{Optical fiber twist sensor with two orthogonally oriented mechanically induced long-period grating sections}},\ }\href {https://doi.org/10.1109/LPT.2004.840073} {\bibfield  {journal} {\bibinfo  {journal} {IEEE Photonics Technology Letters}\ }\textbf {\bibinfo {volume} {17}},\ \bibinfo {pages} {453} (\bibinfo {year} {2005})}\BibitemShut {NoStop}%
\bibitem [{\citenamefont {Wang}\ \emph {et~al.}(2015)\citenamefont {Wang}, \citenamefont {Huang},\ and\ \citenamefont {Wang}}]{Wang2015}%
  \BibitemOpen
  \bibfield  {author} {\bibinfo {author} {\bibfnamefont {Y.}~\bibnamefont {Wang}}, \bibinfo {author} {\bibfnamefont {X.}~\bibnamefont {Huang}},\ and\ \bibinfo {author} {\bibfnamefont {M.}~\bibnamefont {Wang}},\ }\bibfield  {title} {\bibinfo {title} {{Temperature Insensitive Birefringent LPG Twist Sensing Based on the Polarization Properties}},\ }\href {https://doi.org/10.1109/LPT.2015.2466171} {\bibfield  {journal} {\bibinfo  {journal} {IEEE Photonics Technology Letters}\ }\textbf {\bibinfo {volume} {27}},\ \bibinfo {pages} {2367} (\bibinfo {year} {2015})}\BibitemShut {NoStop}%
\bibitem [{\citenamefont {Lin}\ \emph {et~al.}(2022)\citenamefont {Lin}, \citenamefont {Lin}, \citenamefont {Li}, \citenamefont {Xu}, \citenamefont {He}, \citenamefont {Ke}, \citenamefont {Tan}, \citenamefont {Han}, \citenamefont {Li}, \citenamefont {Wang}, \citenamefont {Yao}, \citenamefont {Fu}, \citenamefont {Yu},\ and\ \citenamefont {Cai}}]{Lin2022}%
  \BibitemOpen
  \bibfield  {author} {\bibinfo {author} {\bibfnamefont {Z.}~\bibnamefont {Lin}}, \bibinfo {author} {\bibfnamefont {Y.}~\bibnamefont {Lin}}, \bibinfo {author} {\bibfnamefont {H.}~\bibnamefont {Li}}, \bibinfo {author} {\bibfnamefont {M.}~\bibnamefont {Xu}}, \bibinfo {author} {\bibfnamefont {M.}~\bibnamefont {He}}, \bibinfo {author} {\bibfnamefont {W.}~\bibnamefont {Ke}}, \bibinfo {author} {\bibfnamefont {H.}~\bibnamefont {Tan}}, \bibinfo {author} {\bibfnamefont {Y.}~\bibnamefont {Han}}, \bibinfo {author} {\bibfnamefont {Z.}~\bibnamefont {Li}}, \bibinfo {author} {\bibfnamefont {D.}~\bibnamefont {Wang}}, \bibinfo {author} {\bibfnamefont {X.~S.}\ \bibnamefont {Yao}}, \bibinfo {author} {\bibfnamefont {S.}~\bibnamefont {Fu}}, \bibinfo {author} {\bibfnamefont {S.}~\bibnamefont {Yu}},\ and\ \bibinfo {author} {\bibfnamefont {X.}~\bibnamefont {Cai}},\ }\bibfield  {title} {\bibinfo {title} {{High-performance polarization management devices based on thin-film lithium niobate}},\ }\href
  {https://doi.org/10.1038/s41377-022-00779-8} {\bibfield  {journal} {\bibinfo  {journal} {Light: Science and Applications}\ }\textbf {\bibinfo {volume} {11}},\ \bibinfo {pages} {93} (\bibinfo {year} {2022})}\BibitemShut {NoStop}%
\bibitem [{\citenamefont {Sun}\ \emph {et~al.}(2022)\citenamefont {Sun}, \citenamefont {Xu}, \citenamefont {Li}, \citenamefont {Liu}, \citenamefont {Zhang}, \citenamefont {Cheng}, \citenamefont {Tao}, \citenamefont {Wang}, \citenamefont {Hu}, \citenamefont {Lu}, \citenamefont {Zhao}, \citenamefont {Nie}, \citenamefont {Zhao}, \citenamefont {Guo},\ and\ \citenamefont {Wen}}]{Sun2022}%
  \BibitemOpen
  \bibfield  {author} {\bibinfo {author} {\bibfnamefont {Y.}~\bibnamefont {Sun}}, \bibinfo {author} {\bibfnamefont {Y.}~\bibnamefont {Xu}}, \bibinfo {author} {\bibfnamefont {H.}~\bibnamefont {Li}}, \bibinfo {author} {\bibfnamefont {Y.}~\bibnamefont {Liu}}, \bibinfo {author} {\bibfnamefont {F.}~\bibnamefont {Zhang}}, \bibinfo {author} {\bibfnamefont {H.}~\bibnamefont {Cheng}}, \bibinfo {author} {\bibfnamefont {S.}~\bibnamefont {Tao}}, \bibinfo {author} {\bibfnamefont {H.}~\bibnamefont {Wang}}, \bibinfo {author} {\bibfnamefont {W.}~\bibnamefont {Hu}}, \bibinfo {author} {\bibfnamefont {Y.}~\bibnamefont {Lu}}, \bibinfo {author} {\bibfnamefont {C.}~\bibnamefont {Zhao}}, \bibinfo {author} {\bibfnamefont {T.}~\bibnamefont {Nie}}, \bibinfo {author} {\bibfnamefont {W.}~\bibnamefont {Zhao}}, \bibinfo {author} {\bibfnamefont {Q.}~\bibnamefont {Guo}},\ and\ \bibinfo {author} {\bibfnamefont {L.}~\bibnamefont {Wen}},\ }\bibfield  {title} {\bibinfo {title} {{Flexible Control of Broadband Polarization in a Spintronic
  Terahertz Emitter Integrated with Liquid Crystal and Metasurface}},\ }\href {https://doi.org/10.1021/acsami.2c04782} {\bibfield  {journal} {\bibinfo  {journal} {ACS Applied Materials and Interfaces}\ }\textbf {\bibinfo {volume} {14}},\ \bibinfo {pages} {32646} (\bibinfo {year} {2022})}\BibitemShut {NoStop}%
\bibitem [{\citenamefont {Park}\ \emph {et~al.}(2025)\citenamefont {Park}, \citenamefont {Kim}, \citenamefont {Park},\ and\ \citenamefont {Yoon}}]{Park2025}%
  \BibitemOpen
  \bibfield  {author} {\bibinfo {author} {\bibfnamefont {H.}~\bibnamefont {Park}}, \bibinfo {author} {\bibfnamefont {J.}~\bibnamefont {Kim}}, \bibinfo {author} {\bibfnamefont {G.}~\bibnamefont {Park}},\ and\ \bibinfo {author} {\bibfnamefont {D.~K.}\ \bibnamefont {Yoon}},\ }\bibfield  {title} {\bibinfo {title} {{Enhancing Color Gamut with Wavelength‐Selective Polarization Modulation in Chiral Liquid Crystals}},\ }\href {https://doi.org/10.1002/adom.202402228} {\bibfield  {journal} {\bibinfo  {journal} {Advanced Optical Materials}\ }\textbf {\bibinfo {volume} {13}},\ \bibinfo {pages} {1} (\bibinfo {year} {2025})}\BibitemShut {NoStop}%
\bibitem [{\citenamefont {Zhang}\ \emph {et~al.}(2006)\citenamefont {Zhang}, \citenamefont {Yang}, \citenamefont {Li}, \citenamefont {Yan}, \citenamefont {Yin}, \citenamefont {Gu},\ and\ \citenamefont {Jin}}]{Zhang2006}%
  \BibitemOpen
  \bibfield  {author} {\bibinfo {author} {\bibfnamefont {Y.}~\bibnamefont {Zhang}}, \bibinfo {author} {\bibfnamefont {C.}~\bibnamefont {Yang}}, \bibinfo {author} {\bibfnamefont {S.}~\bibnamefont {Li}}, \bibinfo {author} {\bibfnamefont {H.}~\bibnamefont {Yan}}, \bibinfo {author} {\bibfnamefont {J.}~\bibnamefont {Yin}}, \bibinfo {author} {\bibfnamefont {C.}~\bibnamefont {Gu}},\ and\ \bibinfo {author} {\bibfnamefont {G.}~\bibnamefont {Jin}},\ }\bibfield  {title} {\bibinfo {title} {{Complete polarization controller based on magneto-optic crystals and fixed quarter wave plates}},\ }\href {https://doi.org/10.1364/OE.14.003484} {\bibfield  {journal} {\bibinfo  {journal} {Optics Express}\ }\textbf {\bibinfo {volume} {14}},\ \bibinfo {pages} {3484} (\bibinfo {year} {2006})}\BibitemShut {NoStop}%
\bibitem [{\citenamefont {Stoyanova}\ \emph {et~al.}(2020)\citenamefont {Stoyanova}, \citenamefont {Ivanov},\ and\ \citenamefont {Rangelov}}]{Stoyanova2020}%
  \BibitemOpen
  \bibfield  {author} {\bibinfo {author} {\bibfnamefont {E.}~\bibnamefont {Stoyanova}}, \bibinfo {author} {\bibfnamefont {S.}~\bibnamefont {Ivanov}},\ and\ \bibinfo {author} {\bibfnamefont {A.}~\bibnamefont {Rangelov}},\ }\bibfield  {title} {\bibinfo {title} {{Arbitrary polarization control by magnetic field variation}},\ }\href {https://doi.org/10.1364/AO.404150} {\bibfield  {journal} {\bibinfo  {journal} {Applied Optics}\ }\textbf {\bibinfo {volume} {59}},\ \bibinfo {pages} {10224} (\bibinfo {year} {2020})},\ \Eprint {https://arxiv.org/abs/2011.07834} {2011.07834} \BibitemShut {NoStop}%
\bibitem [{\citenamefont {Li}\ \emph {et~al.}(2007)\citenamefont {Li}, \citenamefont {Wu}, \citenamefont {Dong},\ and\ \citenamefont {Shum}}]{Li2007}%
  \BibitemOpen
  \bibfield  {author} {\bibinfo {author} {\bibfnamefont {Z.}~\bibnamefont {Li}}, \bibinfo {author} {\bibfnamefont {C.}~\bibnamefont {Wu}}, \bibinfo {author} {\bibfnamefont {H.}~\bibnamefont {Dong}},\ and\ \bibinfo {author} {\bibfnamefont {P.}~\bibnamefont {Shum}},\ }\bibfield  {title} {\bibinfo {title} {{Cascaded dynamic eigenstates of polarization analysis for piezoelectric polarization control}},\ }\href {https://doi.org/10.1364/OL.32.002900} {\bibfield  {journal} {\bibinfo  {journal} {Optics Letters}\ }\textbf {\bibinfo {volume} {32}},\ \bibinfo {pages} {2900} (\bibinfo {year} {2007})}\BibitemShut {NoStop}%
\bibitem [{\citenamefont {Meng}\ \emph {et~al.}(2022)\citenamefont {Meng}, \citenamefont {Thrane}, \citenamefont {Ding},\ and\ \citenamefont {Bozhevolnyi}}]{Meng2022}%
  \BibitemOpen
  \bibfield  {author} {\bibinfo {author} {\bibfnamefont {C.}~\bibnamefont {Meng}}, \bibinfo {author} {\bibfnamefont {P.~C.~V.}\ \bibnamefont {Thrane}}, \bibinfo {author} {\bibfnamefont {F.}~\bibnamefont {Ding}},\ and\ \bibinfo {author} {\bibfnamefont {S.~I.}\ \bibnamefont {Bozhevolnyi}},\ }\bibfield  {title} {\bibinfo {title} {{Full-range birefringence control with piezoelectric MEMS-based metasurfaces}},\ }\href {https://doi.org/10.1038/s41467-022-29798-0} {\bibfield  {journal} {\bibinfo  {journal} {Nature Communications}\ }\textbf {\bibinfo {volume} {13}},\ \bibinfo {pages} {2071} (\bibinfo {year} {2022})}\BibitemShut {NoStop}%
\bibitem [{\citenamefont {Awad}\ \emph {et~al.}(2024)\citenamefont {Awad}, \citenamefont {Usmani}, \citenamefont {Virgillito}, \citenamefont {Bratovich}, \citenamefont {Proietti}, \citenamefont {Straullu}, \citenamefont {Aquilino}, \citenamefont {Pastorelli},\ and\ \citenamefont {Curri}}]{Awad2024}%
  \BibitemOpen
  \bibfield  {author} {\bibinfo {author} {\bibfnamefont {H.}~\bibnamefont {Awad}}, \bibinfo {author} {\bibfnamefont {F.}~\bibnamefont {Usmani}}, \bibinfo {author} {\bibfnamefont {E.}~\bibnamefont {Virgillito}}, \bibinfo {author} {\bibfnamefont {R.}~\bibnamefont {Bratovich}}, \bibinfo {author} {\bibfnamefont {R.}~\bibnamefont {Proietti}}, \bibinfo {author} {\bibfnamefont {S.}~\bibnamefont {Straullu}}, \bibinfo {author} {\bibfnamefont {F.}~\bibnamefont {Aquilino}}, \bibinfo {author} {\bibfnamefont {R.}~\bibnamefont {Pastorelli}},\ and\ \bibinfo {author} {\bibfnamefont {V.}~\bibnamefont {Curri}},\ }\bibfield  {title} {\bibinfo {title} {{Environmental Surveillance through Machine Learning-Empowered Utilization of Optical Networks}},\ }\href {https://doi.org/10.3390/s24103041} {\bibfield  {journal} {\bibinfo  {journal} {Sensors}\ }\textbf {\bibinfo {volume} {24}},\ \bibinfo {pages} {3041} (\bibinfo {year} {2024})}\BibitemShut {NoStop}%
\bibitem [{\citenamefont {Usmani}\ \emph {et~al.}(2025)\citenamefont {Usmani}, \citenamefont {D'Amico}, \citenamefont {Straullu}, \citenamefont {Aquilino}, \citenamefont {Bratovich}, \citenamefont {Virgillito},\ and\ \citenamefont {Curri}}]{Usmani2025}%
  \BibitemOpen
  \bibfield  {author} {\bibinfo {author} {\bibfnamefont {F.}~\bibnamefont {Usmani}}, \bibinfo {author} {\bibfnamefont {A.}~\bibnamefont {D'Amico}}, \bibinfo {author} {\bibfnamefont {S.}~\bibnamefont {Straullu}}, \bibinfo {author} {\bibfnamefont {F.}~\bibnamefont {Aquilino}}, \bibinfo {author} {\bibfnamefont {R.}~\bibnamefont {Bratovich}}, \bibinfo {author} {\bibfnamefont {E.}~\bibnamefont {Virgillito}},\ and\ \bibinfo {author} {\bibfnamefont {V.}~\bibnamefont {Curri}},\ }\bibfield  {title} {\bibinfo {title} {{A Smart Sensing Grid for Road Traffic Detection Using Terrestrial Optical Networks and Attention-Enhanced Bi-LSTM}},\ }\href {https://doi.org/10.1109/JLT.2025.3543180} {\bibfield  {journal} {\bibinfo  {journal} {Journal of Lightwave Technology}\ }\textbf {\bibinfo {volume} {PP}},\ \bibinfo {pages} {1} (\bibinfo {year} {2025})}\BibitemShut {NoStop}%
\bibitem [{\citenamefont {Djordjevic}(2022)}]{Djordjevic2022}%
  \BibitemOpen
  \bibfield  {author} {\bibinfo {author} {\bibfnamefont {I.~B.}\ \bibnamefont {Djordjevic}},\ }\bibfield  {title} {\bibinfo {title} {{Quantum sensing and quantum radars}},\ }in\ \href {https://doi.org/10.1016/B978-0-12-822942-2.00007-8} {\emph {\bibinfo {booktitle} {Quantum Communication, Quantum Networks, and Quantum Sensing}}},\ Vol.~\bibinfo {volume} {2}\ (\bibinfo  {publisher} {Elsevier},\ \bibinfo {year} {2022})\ pp.\ \bibinfo {pages} {455--489}\BibitemShut {NoStop}%
\bibitem [{\citenamefont {Lu}\ \emph {et~al.}(2015)\citenamefont {Lu}, \citenamefont {Yu},\ and\ \citenamefont {Oh}}]{Lu2015}%
  \BibitemOpen
  \bibfield  {author} {\bibinfo {author} {\bibfnamefont {X.-m.}\ \bibnamefont {Lu}}, \bibinfo {author} {\bibfnamefont {S.}~\bibnamefont {Yu}},\ and\ \bibinfo {author} {\bibfnamefont {C.~H.}\ \bibnamefont {Oh}},\ }\bibfield  {title} {\bibinfo {title} {{Robust quantum metrological schemes based on protection of quantum Fisher information}},\ }\href {https://doi.org/10.1038/ncomms8282} {\bibfield  {journal} {\bibinfo  {journal} {Nature Communications}\ }\textbf {\bibinfo {volume} {6}},\ \bibinfo {pages} {7282} (\bibinfo {year} {2015})}\BibitemShut {NoStop}%
\bibitem [{\citenamefont {Du}\ \emph {et~al.}(2025)\citenamefont {Du}, \citenamefont {Liu}, \citenamefont {Fadel}, \citenamefont {Vitagliano},\ and\ \citenamefont {He}}]{Du2025}%
  \BibitemOpen
  \bibfield  {author} {\bibinfo {author} {\bibfnamefont {S.}~\bibnamefont {Du}}, \bibinfo {author} {\bibfnamefont {S.}~\bibnamefont {Liu}}, \bibinfo {author} {\bibfnamefont {M.}~\bibnamefont {Fadel}}, \bibinfo {author} {\bibfnamefont {G.}~\bibnamefont {Vitagliano}},\ and\ \bibinfo {author} {\bibfnamefont {Q.}~\bibnamefont {He}},\ }\bibfield  {title} {\bibinfo {title} {{Quantifying entanglement dimensionality from the quantum Fisher information matrix}},\ }\href {http://arxiv.org/abs/2501.14595} {\ ,\ \bibinfo {pages} {1} (\bibinfo {year} {2025})},\ \Eprint {https://arxiv.org/abs/2501.14595} {arXiv:2501.14595} \BibitemShut {NoStop}%
\bibitem [{\citenamefont {Hannun}\ \emph {et~al.}(2022)\citenamefont {Hannun}, \citenamefont {Guo},\ and\ \citenamefont {van~der Maaten}}]{Hannun2022}%
  \BibitemOpen
  \bibfield  {author} {\bibinfo {author} {\bibfnamefont {A.}~\bibnamefont {Hannun}}, \bibinfo {author} {\bibfnamefont {C.}~\bibnamefont {Guo}},\ and\ \bibinfo {author} {\bibfnamefont {L.}~\bibnamefont {van~der Maaten}},\ }\bibfield  {title} {\bibinfo {title} {{Measuring Data Leakage in Machine-Learning Models with Fisher Information}},\ }in\ \href {https://doi.org/10.24963/ijcai.2022/736} {\emph {\bibinfo {booktitle} {Proceedings of the Thirty-First International Joint Conference on Artificial Intelligence}}},\ \bibinfo {series and number} {\bibinfo {number} {Uai}}\ (\bibinfo  {publisher} {International Joint Conferences on Artificial Intelligence Organization},\ \bibinfo {address} {California},\ \bibinfo {year} {2022})\ pp.\ \bibinfo {pages} {5284--5288}\BibitemShut {NoStop}%
\bibitem [{\citenamefont {Bouchet}\ \emph {et~al.}(2021)\citenamefont {Bouchet}, \citenamefont {Rotter},\ and\ \citenamefont {Mosk}}]{Bouchet2021}%
  \BibitemOpen
  \bibfield  {author} {\bibinfo {author} {\bibfnamefont {D.}~\bibnamefont {Bouchet}}, \bibinfo {author} {\bibfnamefont {S.}~\bibnamefont {Rotter}},\ and\ \bibinfo {author} {\bibfnamefont {A.~P.}\ \bibnamefont {Mosk}},\ }\bibfield  {title} {\bibinfo {title} {{Maximum information states for coherent scattering measurements}},\ }\bibfield  {journal} {\bibinfo  {journal} {Nature Physics}\ }\textbf {\bibinfo {volume} {17}},\ \href {https://doi.org/10.1038/s41567-020-01137-4} {10.1038/s41567-020-01137-4} (\bibinfo {year} {2021})\BibitemShut {NoStop}%
\bibitem [{\citenamefont {Guti{\'{e}}rrez-Cuevas}\ \emph {et~al.}(2024)\citenamefont {Guti{\'{e}}rrez-Cuevas}, \citenamefont {Bouchet}, \citenamefont {de~Rosny},\ and\ \citenamefont {Popoff}}]{Gutierrez-Cuevas2024}%
  \BibitemOpen
  \bibfield  {author} {\bibinfo {author} {\bibfnamefont {R.}~\bibnamefont {Guti{\'{e}}rrez-Cuevas}}, \bibinfo {author} {\bibfnamefont {D.}~\bibnamefont {Bouchet}}, \bibinfo {author} {\bibfnamefont {J.}~\bibnamefont {de~Rosny}},\ and\ \bibinfo {author} {\bibfnamefont {S.~M.}\ \bibnamefont {Popoff}},\ }\bibfield  {title} {\bibinfo {title} {{Reaching the precision limit with tensor-based wavefront shaping}},\ }\bibfield  {journal} {\bibinfo  {journal} {Nature Communications}\ }\textbf {\bibinfo {volume} {15}},\ \href {https://doi.org/10.1038/s41467-024-50513-8} {10.1038/s41467-024-50513-8} (\bibinfo {year} {2024})\BibitemShut {NoStop}%
\bibitem [{\citenamefont {Shen}\ and\ \citenamefont {Win}(2007)}]{Shen2007}%
  \BibitemOpen
  \bibfield  {author} {\bibinfo {author} {\bibfnamefont {Y.}~\bibnamefont {Shen}}\ and\ \bibinfo {author} {\bibfnamefont {M.~Z.}\ \bibnamefont {Win}},\ }\bibfield  {title} {\bibinfo {title} {{Fundamental Limits of Wideband Localization Accuracy via Fisher Information}},\ }in\ \href {https://doi.org/10.1109/WCNC.2007.564} {\emph {\bibinfo {booktitle} {2007 IEEE Wireless Communications and Networking Conference}}}\ (\bibinfo  {publisher} {IEEE},\ \bibinfo {year} {2007})\ pp.\ \bibinfo {pages} {3046--3051}\BibitemShut {NoStop}%
\bibitem [{\citenamefont {Barnes}\ \emph {et~al.}(2019)\citenamefont {Barnes}, \citenamefont {Han},\ and\ \citenamefont {Ozgur}}]{Barnes2019}%
  \BibitemOpen
  \bibfield  {author} {\bibinfo {author} {\bibfnamefont {L.~P.}\ \bibnamefont {Barnes}}, \bibinfo {author} {\bibfnamefont {Y.}~\bibnamefont {Han}},\ and\ \bibinfo {author} {\bibfnamefont {A.}~\bibnamefont {Ozgur}},\ }\bibfield  {title} {\bibinfo {title} {{Fisher Information for Distributed Estimation under a Blackboard Communication Protocol}},\ }in\ \href {https://doi.org/10.1109/ISIT.2019.8849821} {\emph {\bibinfo {booktitle} {2019 IEEE International Symposium on Information Theory (ISIT)}}}\ (\bibinfo  {publisher} {IEEE},\ \bibinfo {year} {2019})\ pp.\ \bibinfo {pages} {2704--2708}\BibitemShut {NoStop}%
\bibitem [{\citenamefont {Lee}\ \emph {et~al.}(2022)\citenamefont {Lee}, \citenamefont {Lee}, \citenamefont {Joo},\ and\ \citenamefont {Seok}}]{Lee2022}%
  \BibitemOpen
  \bibfield  {author} {\bibinfo {author} {\bibfnamefont {C.-W.}\ \bibnamefont {Lee}}, \bibinfo {author} {\bibfnamefont {J.~H.}\ \bibnamefont {Lee}}, \bibinfo {author} {\bibfnamefont {J.}~\bibnamefont {Joo}},\ and\ \bibinfo {author} {\bibfnamefont {H.}~\bibnamefont {Seok}},\ }\bibfield  {title} {\bibinfo {title} {{Quantum fisher information of an optomechanical force sensor driven by a squeezed vacuum field}},\ }\href {https://doi.org/10.1364/OE.456731} {\bibfield  {journal} {\bibinfo  {journal} {Optics Express}\ }\textbf {\bibinfo {volume} {30}},\ \bibinfo {pages} {25249} (\bibinfo {year} {2022})}\BibitemShut {NoStop}%
\bibitem [{\citenamefont {Peng}\ \emph {et~al.}(2023)\citenamefont {Peng}, \citenamefont {Qin}, \citenamefont {Zhang},\ and\ \citenamefont {Zhao}}]{Peng2023}%
  \BibitemOpen
  \bibfield  {author} {\bibinfo {author} {\bibfnamefont {Y.}~\bibnamefont {Peng}}, \bibinfo {author} {\bibfnamefont {S.}~\bibnamefont {Qin}}, \bibinfo {author} {\bibfnamefont {S.}~\bibnamefont {Zhang}},\ and\ \bibinfo {author} {\bibfnamefont {Y.}~\bibnamefont {Zhao}},\ }\bibfield  {title} {\bibinfo {title} {{Optical fiber quantum temperature sensing based on single photon interferometer}},\ }\href {https://doi.org/10.1016/j.optlaseng.2023.107611} {\bibfield  {journal} {\bibinfo  {journal} {Optics and Lasers in Engineering}\ }\textbf {\bibinfo {volume} {167}},\ \bibinfo {pages} {107611} (\bibinfo {year} {2023})}\BibitemShut {NoStop}%
\bibitem [{\citenamefont {Gao}(2021)}]{Gao2021}%
  \BibitemOpen
  \bibfield  {author} {\bibinfo {author} {\bibfnamefont {W.}~\bibnamefont {Gao}},\ }\bibfield  {title} {\bibinfo {title} {{Mueller matrix decomposition methods for tissue polarization tomography}},\ }\href {https://doi.org/10.1016/j.optlaseng.2021.106735} {\bibfield  {journal} {\bibinfo  {journal} {Optics and Lasers in Engineering}\ }\textbf {\bibinfo {volume} {147}},\ \bibinfo {pages} {106735} (\bibinfo {year} {2021})}\BibitemShut {NoStop}%
\bibitem [{\citenamefont {Popoff}\ \emph {et~al.}(2010)\citenamefont {Popoff}, \citenamefont {Lerosey}, \citenamefont {Carminati}, \citenamefont {Fink}, \citenamefont {Boccara},\ and\ \citenamefont {Gigan}}]{Popoff2010}%
  \BibitemOpen
  \bibfield  {author} {\bibinfo {author} {\bibfnamefont {S.~M.}\ \bibnamefont {Popoff}}, \bibinfo {author} {\bibfnamefont {G.}~\bibnamefont {Lerosey}}, \bibinfo {author} {\bibfnamefont {R.}~\bibnamefont {Carminati}}, \bibinfo {author} {\bibfnamefont {M.}~\bibnamefont {Fink}}, \bibinfo {author} {\bibfnamefont {A.~C.}\ \bibnamefont {Boccara}},\ and\ \bibinfo {author} {\bibfnamefont {S.}~\bibnamefont {Gigan}},\ }\bibfield  {title} {\bibinfo {title} {{Measuring the Transmission Matrix in Optics: An Approach to the Study and Control of Light Propagation in Disordered Media}},\ }\href {https://doi.org/10.1103/PhysRevLett.104.100601} {\bibfield  {journal} {\bibinfo  {journal} {Physical Review Letters}\ }\textbf {\bibinfo {volume} {104}},\ \bibinfo {pages} {100601} (\bibinfo {year} {2010})}\BibitemShut {NoStop}%
\end{thebibliography}%

\end{document}